%% file: main.tex
\documentclass[lettersize,journal]{IEEEtran}
\usepackage{enumitem}
\usepackage{amsmath,amsfonts}
\usepackage{algorithmic}
\usepackage{algorithm}
\usepackage{array}
\usepackage[caption=false,font=normalsize,labelfont=sf,textfont=sf]{subfig}
\usepackage{textcomp}
\usepackage{stfloats}
\usepackage{url}
\usepackage{verbatim}
\usepackage{graphicx}
\usepackage{cite}
\usepackage{tikz}
\usepackage{amsmath}
\usepackage{listings}
\usepackage{xcolor}
\usepackage{listings}
\usepackage{booktabs}

\usepackage{makecell}
\usepackage{hyperref}
\usepackage[utf8]{inputenc}

\lstdefinelanguage{Solidity}{
    keywords={contract, library, interface, function, modifier, event, struct, enum, mapping, address, uint, int, bool, string, bytes, memory, storage, calldata, public, private, internal, external, pure, view, payable, constant, immutable, override, virtual, abstract, constructor, fallback, receive, revert, require, assert, selfdestruct, keccak256, sha256, ripemd160, ecrecover, addmod, mulmod, blockhash, gasleft, block, msg, tx, abi, type, using, for, new, delete, if, else, while, do, for, break, continue, return, throw, emit, pragma, import, from, as, is},
    keywordstyle=\color{blue}\bfseries,
    ndkeywords={this, super, true, false, nil},
    ndkeywordstyle=\color{darkgray}\bfseries,
    identifierstyle=\color{black},
    comment=[l]{//},
    morecomment=[s]{/*}{*/},
    commentstyle=\color{green!50!black}\ttfamily,
    stringstyle=\color{red}\ttfamily,
    morestring=[b]',
    morestring=[b]",
    sensitive=true,
    breaklines=true,  
    breakatwhitespace=true,  
}

\usepackage[framemethod=TikZ]{mdframed}

\usepackage[most]{tcolorbox}

\newtcolorbox{insightbox}[1][]{
  enhanced,
  breakable,                
  colback=gray!10,          
  colframe=gray!60,         
  boxrule=0pt,              
  leftrule=2pt,             
  sharp corners,            
  boxsep=0pt,               
  top=8pt,                  
  bottom=8pt,               
  left=12pt,                
  right=8pt,                
  fontupper=\normalsize,    
  before skip=12pt,         
  after skip=12pt,          
  #1
}
\usepackage{tabularx} 

\begin{document}

\title{Who Swept My ETH? Uncovering EIP-7702–Based Ether Redirection Attacks on Ethereum}

\author{Ziwei Li, Tianwei Chen, Jiajing Wu,~\IEEEmembership{Senior Member,~IEEE}, Zigui Jiang,~\IEEEmembership{Member,~IEEE}, \\Yuhong Nan,~\IEEEmembership{Member,~IEEE}, Yaobo Huang, Bowen Song, Zibin Zheng,~\IEEEmembership{Fellow,~IEEE}

\thanks{The research is supported by the National Natural Science Foundation of China (62332004, 62472457, 62372485 and 62502548), and the Ant Group.} 

\thanks{Ziwei Li is with the School of Software Engineering, Sun Yat-sen University, Zhuhai 519082, China, and also with Ant Group, Hangzhou 310023, China.}

\thanks{Tianwei Chen, Jiajing Wu, Zigui Jiang, Yuhong Nan and Zibin Zheng are with the School of Software Engineering, Sun Yat-sen University, Zhuhai 519082, China}

\thanks{Yaobo Huang and Bowen Song  are with Ant Group, Hangzhou 310023, China.}
}




\maketitle

\begin{abstract}
EIP-7702 introduces a new transaction type that allows Externally Owned Accounts (EOAs) on major EVM-compatible chains to delegate execution to arbitrary contract codes temporarily. While enhancing functionality, this mechanism facilitates a novel threat: the \underline{E}IP-7702-Based \underline{E}ther \underline{R}edirection \underline{A}ttack (EERA). In such an attack, a victim is delegated to a malicious sweeper contract, which automatically redirects incoming ETH transfers to an address controlled by the attacker. As a nascent attack paradigm arising from protocol evolution, the underlying mechanisms and security implications of EERA remain systematically unexplored. To bridge this gap, we provide the first formal analysis of the EERA attack model and propose SweeperHunter, a framework that detects these attacks through transaction analysis. By evaluating transactions from the first ten months following the introduction of EIP-7702 on Ethereum mainnet, we have identified 4,668,209 malicious delegations, 20,443 attack transactions and 508 ETH sweeper contracts, forming the first EERA dataset. Finally, we analyze this attack from both attacker and victim perspectives, derive key insights, and propose corresponding mitigation strategies. We reveal the prevalence of EERA, as approximately 70\% of the observed delegations point toward ETH sweeper contracts. Furthermore, we observe a notable lack of effective responses from victims; specifically, over 600,000 accounts failed to revoke or update their malicious delegations promptly, thereby remaining in a state of vulnerability. Our research is expected to facilitate understanding and mitigation of this emerging threat posed by EIP-7702.
\end{abstract}


\input{sections/introduction}

\input{sections/Background}
\input{sections/Threat}
\input{sections/SweeperHunter}

\input{sections/results}

\input{sections/analysis}

\input{sections/discussion}

\input{sections/related_work}
\input{sections/conclusion}
\bibliographystyle{IEEEtran}
\bibliography{main}

\end{document}

%% file: sections/introduction.tex
\section{Introduction}\label{sec: introduction}

\IEEEPARstart{E}{thereum}~\cite{ethereum} supports a wide range of decentralized applications through Turing-complete smart contracts. Its account model distinguishes between Externally Owned Accounts (EOAs), which are controlled by private keys but cannot execute custom logic, and Contract Accounts (CAs), which execute predefined smart contract code. To bridge this gap, the Ethereum community has proposed a series of account abstraction mechanisms~\cite{AA}, including EIP-3074~\cite{EIP3074}, EIP-4337~\cite{EIP4337}, and EIP-7702~\cite{EIP7702}. Among them, EIP-7702, activated in May 2025, allows EOAs to delegate their execution logic to smart contracts, enabling advanced functionalities such as transaction batching, gas sponsorship, and flexible permission management. Driven by extensive support from major wallet providers such as MetaMask~\cite{MetaMask} and Trust Wallet~\cite{Trust}, EIP-7702 has been rapidly adopted across the Ethereum ecosystem. According to our statistics, the cumulative number of EIP-7702 delegations increased from 200,521 in May 2025 to more than 6.68 million by February 2026, representing a 33-fold increase within only ten months. While this rapid adoption greatly improves account flexibility, it also introduces a previously unexplored attack surface.

Shortly after the deployment of EIP-7702, GoPlus~\cite{goplus} reported that one of the most frequently authorized delegation contracts was malicious. Once an EOA authorizes such a delegation contract, any ETH subsequently transferred to the account is automatically redirected to an attacker-controlled address. Unlike conventional cryptocurrency theft, this attack exploits the delegated execution mechanism introduced by EIP-7702. We refer to this emerging threat as the \textbf{EIP-7702--Based Ether Redirection Attack (EERA)}, and call the malicious delegation contracts involved in such attacks \textbf{ETH sweeper contracts}.

\begin{figure}[t]
    \centering
    \includegraphics[width=0.8\linewidth]{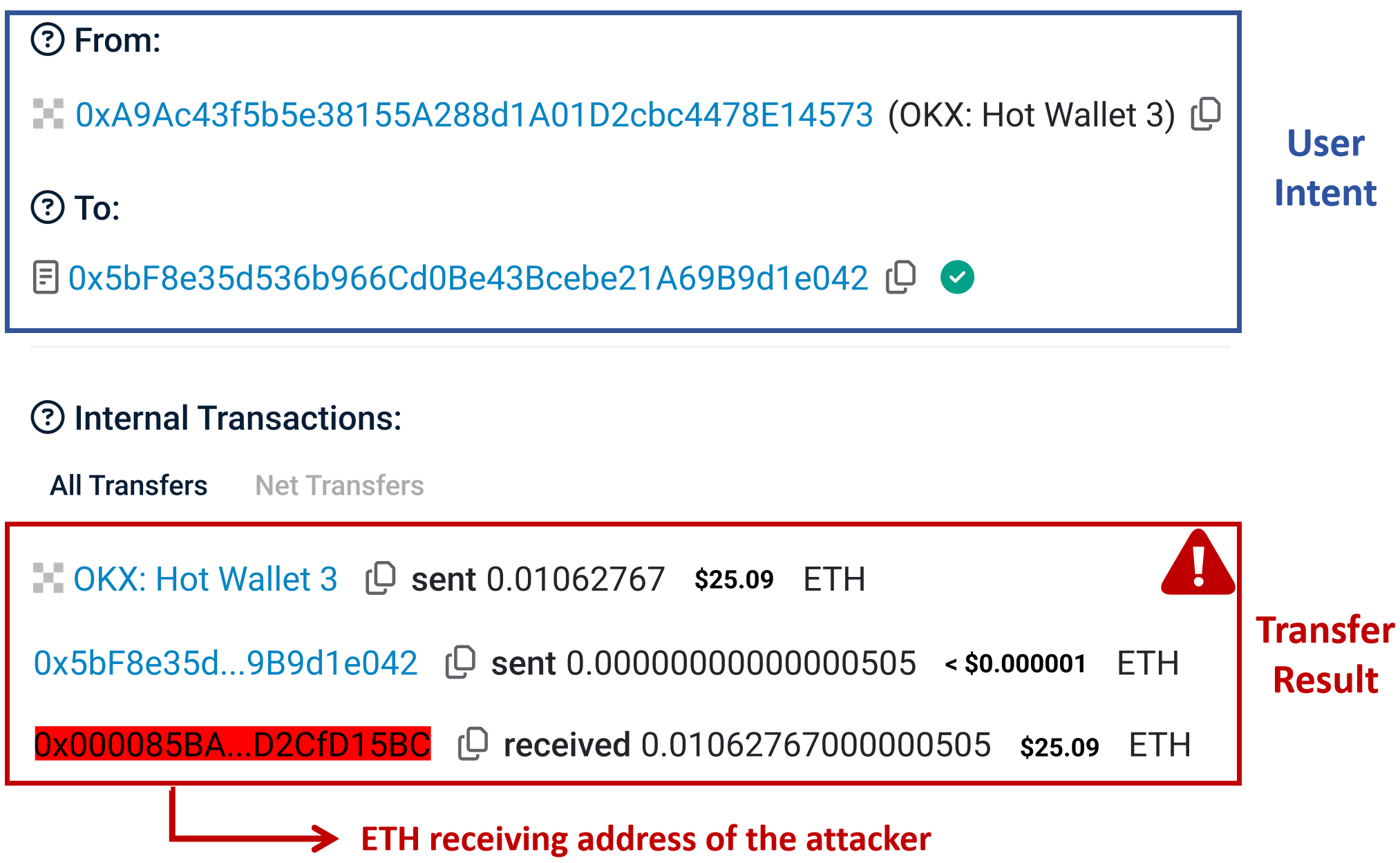}
    \caption{An example of an EERA transaction. After ETH is transferred to a delegated EOA, the delegated contract is automatically executed and redirects both the transferred ETH and the victim's existing balance to an attacker-controlled address. Source: \href{https://etherscan.io/tx/0xec93ee237ae5815b611a88c3032e079013d2bb56a59ede474687ed929a94331f}{Etherscan}.}
    \label{fig:EERA_example}
\end{figure}

\textbf{Rreal-world EERA case}. In this case, an EIP-7702 delegation transaction (\href{https://etherscan.io/tx/0xfc6b71dfe3ef33dd40a16e6ab0c5849a9111d2a66d0f432b381d9ec7e042e388}{\texttt{0xfc6b...e388}}) is first submitted, where the authority address is the victim (\href{https://etherscan.io/address/0x5bF8e35d536b966Cd0Be43Bcebe21A69B9d1e042}{\texttt{0x5bF8...e042}}) and the delegation contract is the ETH sweeper contract. 
 Subsequently, as shown in Fig.~\ref{fig:EERA_example}, a third party transfers ETH to the victim account. Instead of receiving the transferred funds, the delegated contract is automatically executed upon transaction completion and redirects both the transferred ETH and the existing balance of the victim to an attacker-controlled aggregation address. This example demonstrates the complete attack lifecycle, from malicious delegation to automatic asset redirection. Our investigation further reveals that the aggregation address involved in this case has conducted 194 attack transactions, stealing more than \$2,800 worth of ETH from multiple victims.

According to our statistical results, by late February 2026, over \textbf{1.85 million} EOAs have executed more than \textbf{6.68 million} EIP-7702 delegations. Given its role as a foundational standard, the security of EIP-7702 is paramount, as it directly impacts the safety of vast user assets. However, EERA, a novel attack paradigm related to EIP-7702, has yet to receive adequate attention and systematic analysis. To date, efforts have been limited to the disclosure of isolated cases, leaving the underlying mechanisms, the fundamental distinctions from existing attacks, and the overall scale and impact without systematic exploration. The absence of such a comprehensive understanding significantly hinders the ability of the community to effectively mitigate and respond to this emerging threat. 

To bridge this gap, we present the first comprehensive study of EIP-7702--Based Ether Redirection Attacks. Specifically, our contributions are summarized as follows.

\begin{itemize}[leftmargin=*]
    \item  \textbf{Threat Characterization}. To the best of our knowledge, we present the first systematic study of the EERA. Beyond characterizing its execution workflow, we conceptualize EERA as a novel attack paradigm and systematically examine the fundamental shifts it introduces across three dimensions: the passive triggering mechanism, the expanded victim model, and the persistent monetization strategy.
    \item  \textbf{Empirical Dataset}. We present the first large-scale EERA dataset, namely \textit{\textbf{SweeperDataset}}. We analyze in-the-wild transactions from May 2025 to February 2026, confirming over 4.66 million malicious delegations and 20,443 EERA transactions involving 508 ETH sweeper contracts. \textit{\textbf{SweeperDataset}} provides a foundational resource to analyze this new attack vector and develop defense mechanisms.
    \item  \textbf{Ecosystem Analysis}. By analyzing the behaviors of attackers and victims, we reveal the costs and profits of attackers, the implementation methods of delegation contracts, and the destination of ETH. Furthermore, we explain the scale of victims, the characteristics of malicious delegations, and the responses of victims to these malicious delegations.
    \item  \textbf{Practical Impact}. We actively contribute to the security of the community by reporting the labels, findings and mitigation recommendations of this study to GoPlus and Chainabuse. We have received positive feedback from GoPlus, facilitating the identification of EERA threats.
\end{itemize}

%% file: sections/Background.tex
\section{Background}\label{sec:background}

\subsection{Ethereum Accounts}
Ethereum adopts an account-based state model. Each account comprises four core fields: \textit{balance}, \textit{nonce}, \textit{storage root}, and \textit{code hash}. The \textit{nonce} records the number of transactions sent by the account or the number of contract creations to prevent transaction replay. The \textit{balance} represents the amount of Ether held by the account. The \textit{storage root} and \textit{code hash} correspond to the persistent storage state and the deployed bytecode of a contract account, respectively. Based on the presence of contract code, Ethereum accounts are classified into two categories: Externally Owned Accounts (EOAs) and Contract Accounts. An EOA is controlled by a private key and can initiate transactions directly. In contrast, a contract account is driven by on-chain code and executes only when called. The creation and execution of contract accounts depend on external accounts; an EOA can deploy new contracts by initiating transactions or trigger the execution of code by sending transactions to existing contracts.

\subsection{Ethereum Transactions}
An Ethereum transaction serves as a cryptographically signed instruction initiated by an EOA to trigger state transitions within the network. Each transaction comprises essential fields, including the sender address, the recipient address, the transfer amount (\texttt{value}), input data (\texttt{calldata}), gas limit, and fee-related parameters. A transaction is interpreted as a contract creation if the recipient field is null; otherwise, it constitutes a message call targeting either another EOA or a smart contract. When the target is a smart contract, the Ethereum Virtual Machine (EVM) executes the associated bytecode, which may trigger internal calls to other contracts. These internal operations are typically facilitated by instructions such as \texttt{CALL}, \texttt{DELEGATECALL}, and \texttt{STATICCALL}, which enable the transfer of Ether, data, and execution context between accounts. Transaction execution necessitates the consumption of Gas, a unit used to measure computational and storage resource requirements. Users specify a gas limit and fee parameters upon submission, with validators prioritizing transactions that offer higher fees. The final transaction fee, settled in Ether (the native token of Ethereum), is determined by the actual amount of Gas consumed during execution.

%% file: sections/Threat.tex
\section{Threat: EIP-7702 Ether Redirection Attack}\label{sec:Threat}

In this study, we identify a novel attack that exploits the EIP-7702 delegation mechanism.  By analyzing disclosed cases, EIP-7702’s design, and on-chain data, we systematically characterize the underlying principles and operational workflows of this attack.

\subsection{Attack Flow}

\begin{figure}[t]
    \centering
    \includegraphics[width=0.85\linewidth]{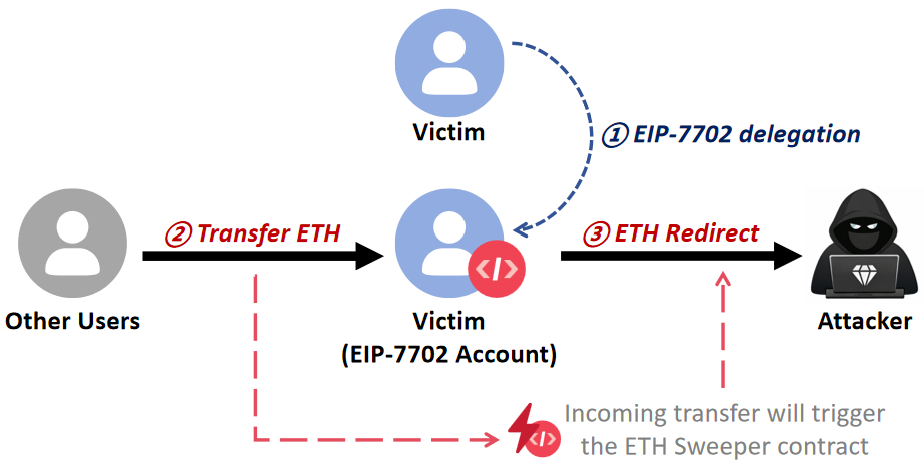}
    \caption{The attack flow of EERA. A third-party account initiates an ETH transfer to the victim that is in a delegation state. This transfer automatically triggers the sweeping logic of the malicious delegation contract, resulting in the redirection of ETH to an attacker-controlled address.}
    \label{fig: attack_process}
\end{figure}

The attack flow is shown in Fig.~\ref{fig: attack_process}, including the following three steps:
\textcircled{1} \textbf{Malicious Delegation:} The attacker first deploys an ETH sweeper contract that redirects incoming ETH to an attacker-controlled address. A delegation transaction is then submitted to delegate the victim account to this contract. It is important to note that any entity can submit this delegation transaction as long as they possess the authorization signature confirmed by the EOA. Since the authorization signature is derived from the private key of the authority, it can be obtained either by inducing the victim to actively sign a malicious authorization or by the attacker performing the delegation directly after compromising the private key of the victim.
\textcircled{2} \textbf{Third-party Transfer:} Under the delegated state, once another user sends ETH to the address of the victim, the victim serves as the \texttt{to\_address} of the transaction. Upon receiving the ETH, the code of the delegation contract is triggered for execution.
\textcircled{3} \textbf{Ether Sweeping:} The pre-defined ETH sweeping logic in the malicious delegation contract executes automatically, transferring the ETH intended for the account of the victim or the entire balance of the account of the victim to a phishing address preset by the attacker, thus completing the Ether redirection. Note that the victim accepts the ETH before the code of the delegation contract is executed; therefore, if the code transfers \texttt{address(this).balance}, it transfers both the ETH received in the current transaction and the original ETH balance of the victim. Overall, by delegating the victim to a malicious contract, the attacker can leverage normal third-party ETH transfer behaviors to continuously and automatically execute redirection code, achieving ETH theft.

\subsection{Key Shifts}
By decoupling account identity from execution logic, EERA introduces a fundamental shift in the on-chain attack paradigm, transforming traditional asset theft across three dimensions.

\textbf{(1)  Shift in the Triggering Mechanism.} Traditional asset-draining attacks typically require a victim to initiate a transaction or invoke a specific function within a malicious contract. In contrast, the execution of an EERA is automated and remains passive from the perspective of the victim account. Following a malicious delegation, the attack is triggered by the transfer of ETH from a third-party account to the authority. This process requires no participation from the victim account and involves no active interaction with the malicious contract, which significantly enhances the stealth of the attack.

\textbf{(2)  Shift in the Victim Model.} In conventional asset theft scenarios, the impact is typically restricted to the direct owners of the compromised accounts. However, the EERA model introduces a shift where the scope of potential victims extends beyond the authorities who sign the malicious delegations. This paradigm involves third-party users who transmit funds to these compromised addresses, as they may remain unaware of the subsequent redirection. For instance, in scenarios such as automated payroll distribution, a sender might inadvertently become an indirect victim if the transferred ETH is immediately siphoned by an attacker. Since these third-party participants do not necessarily possess knowledge of the underlying malicious delegation, the EERA model poses a broader systemic risk where even well-intentioned transfer contribute to the success of the attack.

\textbf{(3) Shift in the Monetization Strategy.} Unlike traditional attacks that extract a fixed amount of assets from the existing balance of a victim account, EERA primarily monetizes future incoming transactions. Once a victim account is delegated to a malicious contract, any subsequent Ether transferred to this account can be automatically redirected to the attacker. As a result, the attacker profit is no longer bounded by the current holdings of the victim, but instead accumulates over time with continuous inflows. This transforms the attack from one-time asset extraction into a persistent monetization mechanism.


%% file: sections/SweeperHunter.tex
\section{Detection of EERA}\label{sec:detection}


Regarding the EERA, we observe that this type of attack simultaneously satisfies two key characteristics: (1) the attack occurs while the victim is in an EIP-7702 delegated state, and the code of the delegation contract is executed during the attack; (2) the execution of the transaction results in the malicious redirection of the ETH of the victim, exhibiting clear asset-sweeping behavior. Based on these understandings, we propose \textit{SweeperHunter}, an EERA detection system based on transaction analysis. Specifically, the process of SweeperHunter consists of three main stages. \textbf{First}, the system collects transaction metadata and execution traces. \textbf{Second}, it identifies and parses EIP-7702 delegation transactions to construct a delegation state mapping table for each \textit{Authority} based on the timeline of the delegation. \textbf{Finally}, SweeperHunter detects EERA attacks by applying a set of heuristic rules.

\subsection{Data Collection}

To construct a dataset for analysis, \textit{SweeperHunter} collects historical transaction data from the Ethereum mainnet using the \textit{BlockchainSpider} tool~\cite{blockchainspider}. 
Given that EIP-7702 was activated in May 2025, our study spans a 10-month period from May 1, 2025, to March 1, 2026, encompassing the entire protocol lifecycle from its inception to the most recent activities. This comprehensive coverage allows us to capture the full trajectory of the EERA threat, providing a longitudinal perspective on its real-world adoption and the associated security risks.
For each transaction, we retrieve the following two types of data, which provides the necessary information to identify EIP-7702 delegation relationships and analyze subsequent asset movements. \textbf{(1) Transaction Metadata:} This includes essential fields such as the transaction hash, block number, timestamp, participating addresses (\texttt{from\_address} and \texttt{to\_address}), and the amount of ETH transferred. \textbf{(2) Execution Traces:} These records detail the internal operations performed by the EVM. Specifically, our system collects the sequence of opcodes (e.g., \texttt{CALL}, \texttt{DELEGATECALL}) and their associated data, such as input parameters and value transfers.

\subsection{Delegation Mapping}

In this step, we construct a delegation mapping table by identifying EIP-7702 delegation transactions and resolving the corresponding \textit{authority}. 

EIP-7702 delegation transactions are identified by the transaction type \texttt{0x4}. Each transaction contains an \texttt{authorizationList}, which records the delegated contract address together with the signature components, but omits the explicit \textit{authority}. Following the EIP-7702 specification~\cite{EIP7702}, we recover the \textit{authority} from the authorization signature. First, we compute the authorization message hash $H$:

\begin{equation}
    H = \text{keccak256}(0x05 \parallel \text{rlp}([chain\_id, address, nonce]))
    \label{equ:hash}
\end{equation}
where $chain\_id$, $address$, and $nonce$ are the values taken directly from the authorization entry. Here, $address$ refers to the address of the delegated contract. The constant \texttt{0x05} is a version identifier, $\parallel$ denotes byte-wise concatenation, and $\text{rlp}$ is the Recursive Length Prefix encoding function~\cite{rlp}.

Second, we use the elliptic curve public key recovery function, \texttt{ecrecover}, with the hash $H$ and the signature components to derive the address of the \textit{authority}:
\begin{equation}
    authority = \texttt{ecrecover}(H, y\_parity, r, s)
    \label{equ:authority}
\end{equation}
where $y\_parity$, $r$, and $s$ are the signature values from the authorization entry. 

Not all delegation transactions become effective because protocol-level validation (e.g., signature, nonce, and chain-ID verification) may fail. We therefore verify each recovered delegation by querying the account code of the \textit{authority}. A valid delegation sets the account code to \texttt{0xef0100} concatenated with the delegated contract address. Delegations that do not satisfy this condition are discarded.

Finally, for each \textit{authority}, we construct a chronological delegation mapping table,

\begin{equation}
[(t_1,C_1),(t_2,C_2),\dots,(t_n,C_n)],
\end{equation}

where $t_i$ denotes the delegation timestamp and $C_i$ denotes the delegated contract (or \texttt{0x0} for revocation). This table is subsequently used to determine the delegation state of an account at any transaction timestamp.

\subsection{Attack Discovery}

\begin{table}[t]
\centering
\caption{Heuristic Rules for EERA Detection}
\label{tab:heuristic_rules}
\begin{tabular}{cll}
\toprule
\textbf{Rule} & \textbf{Constraint} & \textbf{Formal Condition} \\ 
\midrule
R1 & Transfer Intent & $tx.value \neq 0 \land tx.input = 0$ \\
R2 & Transfer Triggered & $trace.transfer.value \neq 0$ \\
R3 & State Verification & $is\_delegated(tx.to, t_{tx}) = \text{True}$ \\
R4 & ETH Redirection & $trace.transfer.from = tx.to$ \\
R5 & ETH Redirection & $trace.transfer.to \notin \{tx.from, tx.to\}$ \\ 
\bottomrule
\multicolumn{3}{l}{
\begin{minipage}{0.95\linewidth}
\smallskip
\footnotesize
$tx.from$ and $tx.to$ denote the initiator and semantic recipient; $tx.value$ and $tx.input$ represent the ETH amount and input data. Regarding the execution trace, $trace.transfer.from$, $trace.transfer.to$, and $trace.transfer.value$ record the sender, recipient, and amount moved during execution. $t_{tx}$ denotes the transaction timestamp. $is\_delegated(address, t)$ returns $\text{True}$ if the address is in an active EIP-7702 delegation state at $t$.
\end{minipage}}
\end{tabular}
\end{table}

We detect EERA by cross-referencing crawled transaction data with our constructed delegation mapping table using a formalized rule-based approach. In an EERA attack, two distinct ETH transfers occur. First, ETH is transferred from a third-party account to the account of the victim. This transfer is reflected in the transaction metadata, such as the \texttt{from}, \texttt{to}, and \texttt{value} fields. Subsequently, an ETH transfer caused by the automatic triggering of a sweeping contract occurs, moving ETH from the account of the victim to an account specified by the attacker. Since this second transfer is a result of contract execution, the behavior is recorded in the transaction trace. Therefore, the core of EERA detection involves analyzing the inconsistency between these two fund transfers to capture the critical behavior of redirection. Based on this understanding, we design the heuristic rules (\textbf{R}) to evaluate transactions, which are shown in Table~\ref{tab:heuristic_rules}. 

Our detection rules are designed as \textbf{sufficient conditions} for EERA; that is, any transaction identified is logically guaranteed to satisfy the semantic definition of an EERA attack. 
Specifically, \textbf{(1) Transfer Intent:} R1 restricts transactions to those with empty calldata and non-zero value, aiming to identify cases where the $tx.to$ address corresponds to the intended ETH recipient rather than a contract used for functional execution. In such transactions, the recipient account may trigger automatic execution of its delegated contract code, which aligns with the execution model assumed in EERA. In contrast, transactions with non-empty calldata explicitly invoke the logic of a target contract, where any ETH transfer is governed by contract semantics rather than account-level delegated execution, and thus falls outside the scope of EERA. 
\textbf{(2) Transfer Triggered:} R2 requires that an ETH transfer is triggered during the execution of the contract code, which serves as the fundamental basis for observing potential redirection behavior within the trace.
\textbf{(3) State Verification:} R3 ensures that the target address of the transaction is in an active delegated state at the time of execution, linking the behavior directly to EIP-7702.
\textbf{(4) ETH Redirection:} R4 and R5 detect the core fund redirection behavior, where ETH sent from a third-party account to a specific recipient is ultimately credited to an unrelated external account.

When the metadata and execution trace of a transaction simultaneously satisfy the aforementioned conditions, we flag it as a suspicious EERA attack transaction. For every identified transaction, \textit{SweeperHunter} further records the transaction metadata, the complete execution trace, and the contract delegated by $tx.to$ at the time of the transaction to support subsequent attack verification and analysis.

%% file: sections/results.tex
\section{Detection Results}\label{sec:results}

\subsection{EIP-7702 Delegations}

\begin{figure}[t]
    \centering
    \includegraphics[width=1\linewidth]{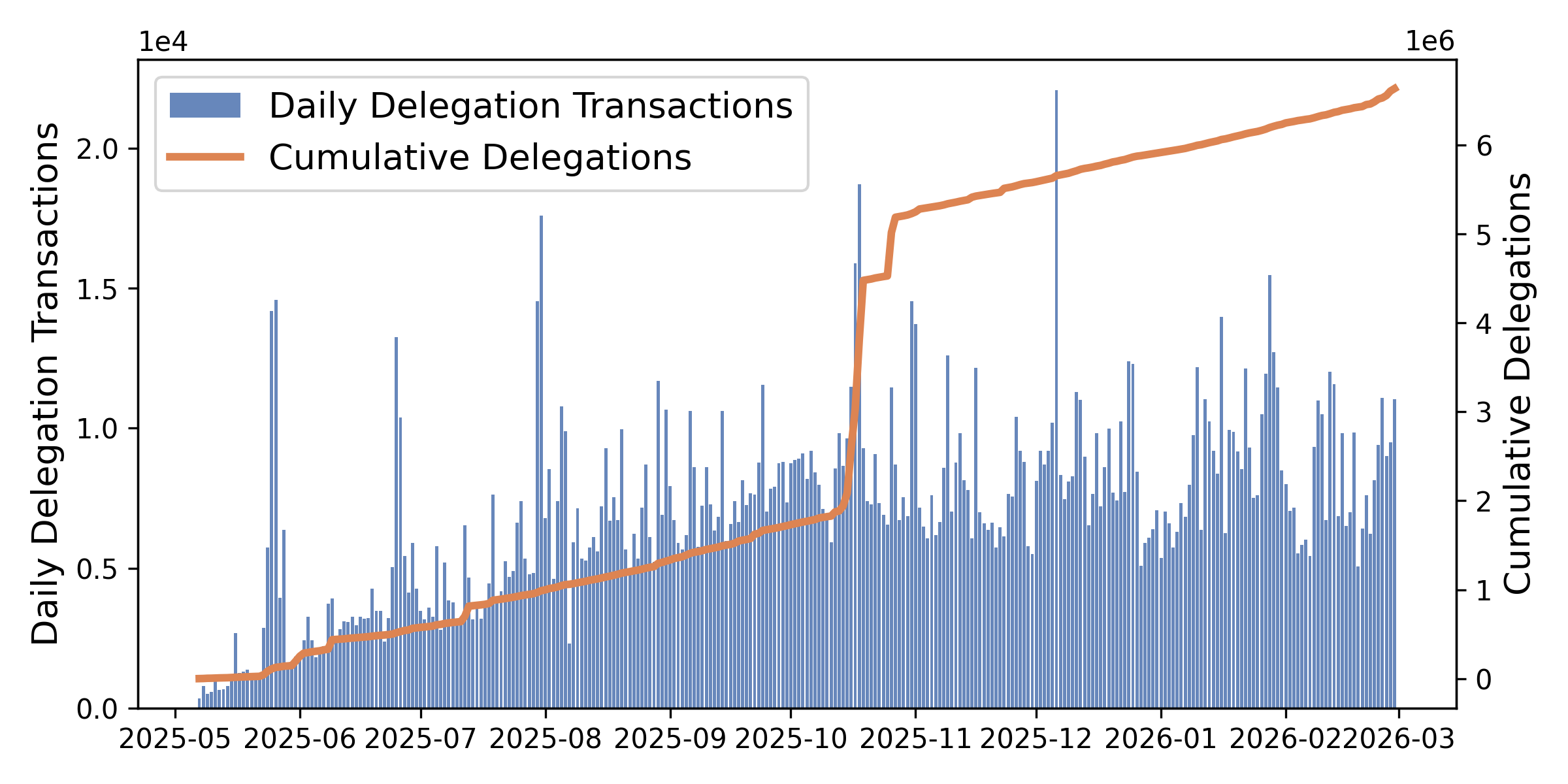}
    \caption{The temporal distribution of EIP-7702 delegations}
    \label{fig: delegation_data}
\end{figure}

\begin{table}[t]
\centering
\setlength{\belowcaptionskip}{10pt} 
\caption{Monthly Volume of EIP-7702 Delegations}
\label{tab:delegation_volume}
\begin{tabular}{ccc}
\toprule
\textbf{Month} & \textbf{\#Valid Transactions} & \textbf{\#Valid Delegations} \\ 
\midrule
2025-05       & 70,293       & 200,521     \\
2025-06      & 117,577      & 374,022     \\
2025-07      & 163,641      & 419,424     \\
2025-08     & 220,148      & 341,684     \\
2025-09    & 236,999      & 407,493     \\
2025-10   & 286,059      & 3,512,603   \\
2025-11   & 237,379     &   353,406 \\
2025-12   & 281,332     &    336,050\\
2026-01   &  289,352    &  329,353   \\
2026-02  &  233,716    &  405,942   \\
\midrule
Total &  2,136,496 & 6,680,498\\
\bottomrule
\end{tabular}
\end{table}

Within the transaction data spanning from May, 2025 to February 2026, a total of 2,308,833 EIP-7702 delegation transactions are detected, encompassing 6,922,781 individual account delegations. Through the authority parsing and delegation result confirmation processes of \textit{SweeperHunter}, it is determined that \textbf{6,680,498} (96.50\%) delegations contained in \textbf{2,136,496} transactions are valid delegations that successfully pass the protocol-level verification.


The temporal distribution of these valid delegations and their corresponding transactions is shown in Fig.~\ref{fig: delegation_data}. Additionally, the monthly statistics for delegation volume are summarized in Table~\ref{tab:delegation_volume}. Between May and September 2025, the daily delegation volume maintains robust linear growth as the EIP-7702 delegation ecosystem expands. Then, a significant surge occurs in mid-October 2025, initiating a period of sustained high activity through March 2026 that drives the cumulative volume to exceed \textbf{6.68 million}.
The spike in October 2025 is structurally characterized by a high frequency of batch delegations, where 286,063 transactions contained 3,512,607 authorizations, averaging 12 delegations per transaction. Subsequent analysis in Section~\ref{sec: Analysis} indicates that malicious delegations account for at least 78.92\% of these authorizations. Furthermore, we find that this sudden surge is partially driven by attackers executing large-scale, automated exploitations of leaked private keys. We cross-referenced our dataset with the repository of compromised Ethereum private keys collected by Shao et al.~\cite{lost_in_the_address}. Strikingly, 4,050 out of the 4,189 (96.7\%) reported leaked addresses were targeted precisely during this October spike, generating 250,520 malicious delegations. And 99.2\% (248,598) delegations were bundled inside batched delegation transactions to maximize gas efficiency.

\subsection{EERA Transactions}

\begin{figure}[t]
    \centering
    \includegraphics[width=1\linewidth]{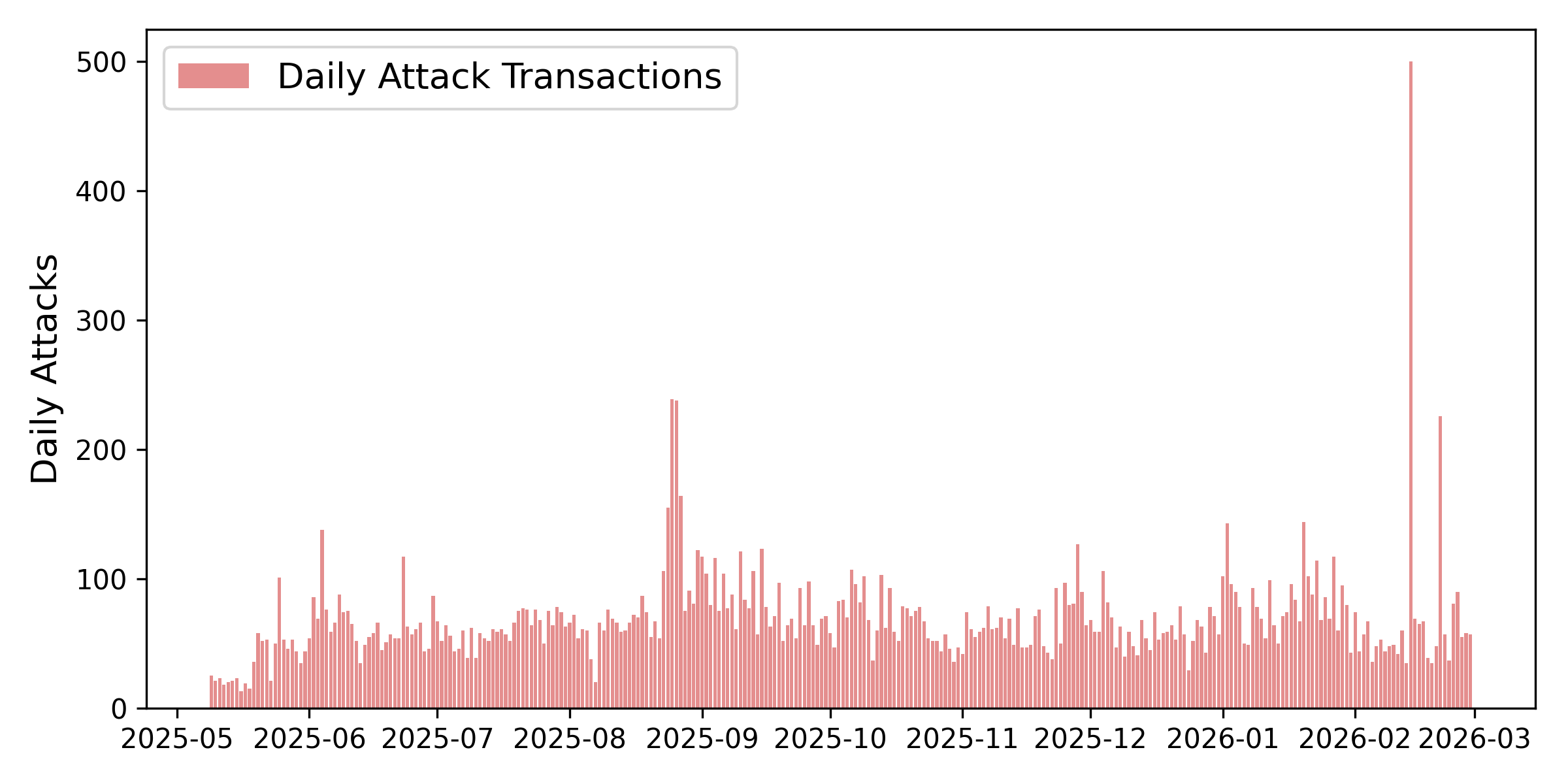}
    \caption{The temporal distribution of these detected attack transactions}
    \label{fig:attack_distribution}
\end{figure}


We have detected a total of 20,443 attack transactions involving 508 delegation contracts. As shown in Fig.~\ref{fig:attack_distribution}, EERA attacks occur consistently throughout the observation period, with a notable peak on February 14, 2026. Further inspection shows that 462 of the 500 attacks on that day have been initiated by a single account (\texttt{0x2840...8057}) within only 9 hours, targeting 286 victim addresses through highly repetitive ETH transfers. The core delegation contracts involved (e.g.,\texttt{0xb59f...dfca}) have been flagged as malicious by Wintermute and Etherscan. Although the campaign has transferred less than 0.0015 ETH in total, every transaction satisfies the complete execution flow of EERA. We conjecture that these transactions reflect a validation campaign, in which the attacker verifies the effectiveness of previously deployed malicious delegations before conducting exploitation.

\textbf{Evaluation of Detection Results.} We evaluate the effectiveness of \textit{SweeperHunter} through a combination of delegation contract analysis and transaction sampling analysis.
\textbf{(1) Contract-level Evaluation.}
We perform a manual analysis of all 508 detected delegation contracts. Specifically, we leverage \textit{Dedaub}~\cite{deadub} to decompile the contract bytecodes, which are then independently examined by three blockchain security experts following a standardized logic-verification. The experts unanimously confirm that all 508 contracts implement malicious ETH sweeping behavior, achieving a consensus with zero ambiguity.
Furthermore, we cross-reference these contracts with external intelligence sources. We find that 42 contracts are explicitly labeled by the security firm \textit{WinterMute}~\cite{wintermute} as EIP-7702-related crime contracts, while none are categorized as benign services (e.g., wallets or infrastructure), indicating no  contradictions against available ground truth.
\textbf{(2) Transaction-level Evaluation.}
Following standard practices~\cite{ndss2025, usenix2023}, we perform stratified random sampling by selecting one transaction for each of the 508 contracts. Three experts independently audit these 508 sampled transactions by correlating \textit{Etherscan} execution traces with the decompiled contract semantics. All sampled transactions are unanimously confirmed as authentic EERA instances, demonstrating the high precision of our methodology. The procedure of manual analysis is detailed in our repository~\footnote{https://github.com/liziwei1/SweeperHunter}. Overall, we do not observe false positives in our detect results. This is mainly because our detection rules jointly constrain delegation state, value transfer, and execution trace consistency, which tightly characterize the semantics of EERA.

\textbf{Adversarial Evasion and Boundary Analysis.}
The rule-based nature introduces specific evasion boundaries and potential false negatives.  \textit{\textbf{On one hand}}, attackers may attempt to evade detection by transferring ETH to a victim through a redirection target address to trigger the execution of a sweeper contract. This strategy bypasses our heuristic rule (R5), which requires \texttt{trace.to} $\neq$ \texttt{tx.from}, because no explicit redirection occurs within the transaction trace. Under this evasion scenario, an attacker must proactively fund the victim and deploy a \textit{this.balance-based} sweeper contract (as introduced in Section~\ref{sec:payload-contract}) to capture the pre-existing assets of the victim as illicit profit. However, we argue that this approach is cost-ineffective. First, the operational overhead increases as the attacker must personally bear the gas fees for the initial transfer; Second, the potential profit is strictly limited to the pre-existing balance of the victim, rather than the more lucrative interception of incoming ETH from third parties. Consequently, such an evasion strategy is economically irrational, as it contradicts the fundamental assumption that on-chain attack behaviors are driven by the pursuit of maximized economic interests. Therefore, we conclude that this specific evasion tactic poses a negligible threat to the robustness of \texttt{SweeperHunter}.
\textit{\textbf{On the other hand}}, while EIP-7702 could be exploited to sweep non-native tokens (e.g., ERC-20), such derivative attacks fall outside the current detection scope of \texttt{SweeperHunter}. Because non-native token transfers necessitate external contract calls rather than the native value-transfer mechanisms of the EVM, the underlying execution traces differ fundamentally from ETH-based EERA. This paper does not aim to exhaustively analyze all threats amplified by EIP-7702; instead, the scope is deliberately restricted to EERA, a critical, pervasive, and previously uncharacterized class of native redirection attacks.

The detection results form the first dataset regarding EERA, which we refer to as \textit{\textbf{SweeperDataset}} and make public.


%% file: sections/analysis.tex
\section{Analysis}\label{sec: Analysis}
Based on \textit{SweeperDataset}, we analyze the EERA from the perspectives of both attackers and victims. Specifically, first, we analyze the costs and profits of attackers, the attack payloads, and the flow of illicit funds. Second, we examine the scale of victims, the characteristics of malicious delegations, and the response of victims following malicious delegations.

\subsection{Attackers}
\subsubsection{\textbf{Cost and Profit}}

The profit of an attacker is defined as the amount of ETH transferred to the attacker-specified account. The associated costs comprise a one-time \textit{contract deployment fee} (which represents a negligible fraction of the total) and \textit{delegation transaction fees}. We clarify two cost-accounting boundary conditions: \textbf{First}, since on-chain data cannot definitively confirm whether the attacker funded a delegation transaction, we count the fee as part of the attacker cost as long as the transaction was not submitted by the victim. Consequently, the statistical results represent the upper bound of the potential cost to avoid underestimation. \textbf{Second}, rather than accounting only for delegation transactions that directly led to successful attacks, we incorporate all delegation fees associated with the detected sweeper contracts. These comprehensive fees reflect the necessary capital investment made during the preparation stage.

\begin{figure}
    \centering
    \includegraphics[width=1\linewidth]{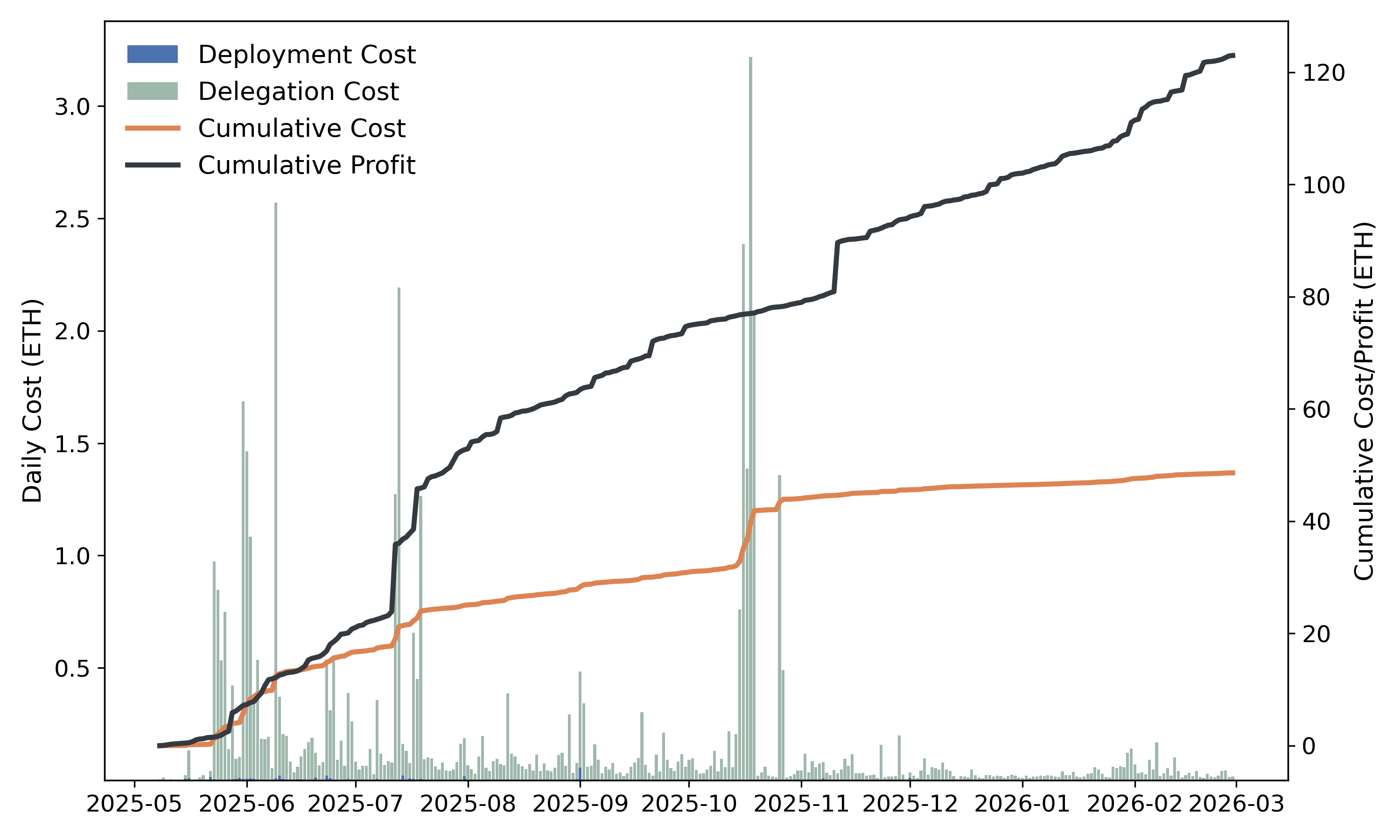}
    \caption{The costs and profits of attackers, where the bar chart represents the contract deployment and delegation costs, the orange line represents cumulative costs, and the black line represents cumulative profits.}
    \label{fig:cost_and_profit}
\end{figure}

Fig.~\ref{fig:cost_and_profit} illustrates the statistical results of the costs and profits of attackers. Specifically, attackers realize a total profit of \textbf{122.98} ETH through 20,443 attack transactions, while incurring costs that include \textbf{48.14} ETH in delegation fees (involving 641,379 delegation transactions) and \textbf{0.47} ETH in contract deployment fees. The economic model of EERA is fundamentally characterized by a \textit{front-loaded investment} coupled with \textit{long-term passive extraction}. Unlike many on-chain attacks that seek immediate exploitation, EERA requires attackers to first absorb substantial gas costs to deploy sweeper contracts and broadcast malicious delegations across a vast pool of potential victims. This initial expenditure creates a latent threat; profitability is not instantaneous but is instead contingent on future ETH transfers to the compromised accounts, resulting in a distinct temporal lag between investment and return. Once established, however, these persistent delegations function as automated traps, enabling the continuous and systematic siphoning of assets over time. Rather than a singular, high-profile exploit, EERA manifests as a cumulative threat that yields steady returns for the attacker while remaining a perennial risk.

\begin{insightbox}
    \textbf{Finding 1:} \textit{Attackers have illicitly seized 122.98 ETH. Their primary overhead is confined to the gas fees of delegation transactions, as the monetization paradigm shifts toward persistent value extraction that depends on subsequent ETH inflows from third-party accounts.}
\end{insightbox}

\subsubsection{\textbf{Attack Payload}}\label{sec:payload-contract} ETH sweeper contracts constitute the core payload of the attack. We therefore further analyze their deployment, delegations, and implementation.

\begin{figure}
    \centering
    \includegraphics[width=1\linewidth]{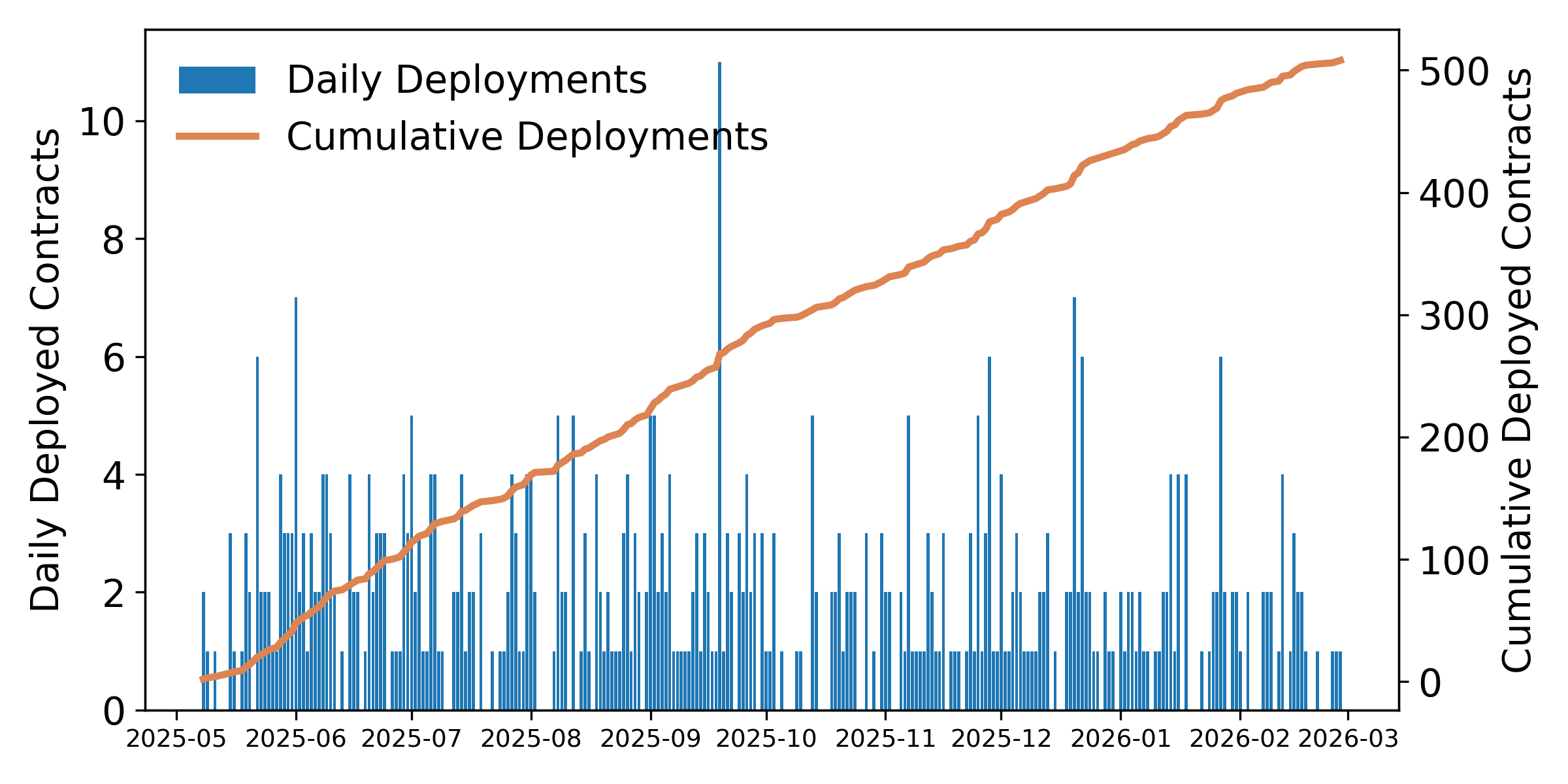}
    \caption{The temporal distribution of ETH sweeper contract deployments}
    \label{fig:contract_deploy}
\end{figure}

\textbf{Contract Deployment.} We utilize the interface provided by \textit{Etherscan} (\texttt{getcontractcreation}) to crawl the deployment information of the 508  malicious delegation contracts, including deployment timestamps, deployer addresses, and corresponding transaction hashes. These contracts correspond to 330 distinct deployer addresses, implying that certain attacker accounts deploy multiple malicious ETH sweeper contracts. The addresses \href{https://etherscan.io/address/0xa6e660ea5ed28a8de90557adfaad254f94ddafea}{\texttt{0xa6e6...afea}} and \href{https://etherscan.io/address/0xcc592ef7c86a7e4d8fa6a725b1244bd80133aa52}{\texttt{0xcc59...aa52}}  have created 8 ETH sweeper contracts, making them the two addresses with the highest number of such contract deployments. Given that the cost of contract deployment is relatively low and remains almost negligible within the overall expenditure of an attacker, the deployment of multiple contracts offers a cost-effective approach to achieve several objectives. \textit{First, risk isolation}. By utilizing independent contracts, attackers can isolate their activities and reduce the risk of large-scale disruption if a single contract is flagged or blacklisted. \textit{Second, customized redirection}. If a change in the destination for funds redirection is required and the previous contract lacks a update function, attackers may deploy a new contract to implement the necessary modifications to the redirection target. 

We further analyze the temporal distribution of these contract deployments, as illustrated in Fig.~\ref{fig:contract_deploy}. Overall, the deployment of contracts exhibits a balanced distribution over time, suggesting that attackers continuously build new delegation contracts to expand their attack payloads. On September 19, the volume of deployed delegation contracts reached a peak of 11. Interestingly, 6 of these contracts were created by the same address \href{https://etherscan.io/address/0x26fddb62cdb05ff4c787c2ec6b0a280fa89e28ce}{\texttt{0x26fd...28ce}}. 
Furthermore, during the manual audit of these contracts, experts observe significant code reuse: most contracts show high similarity in their decompiled code, differing only in key parameters such as the redirection target address. 
These findings suggest that attackers likely adopt templated or batch-generation methods to deploy attack components, providing important insights for future detection methods of ETH sweeper contracts based on code similarity.

\begin{figure}
    \centering
    \includegraphics[width=1\linewidth]{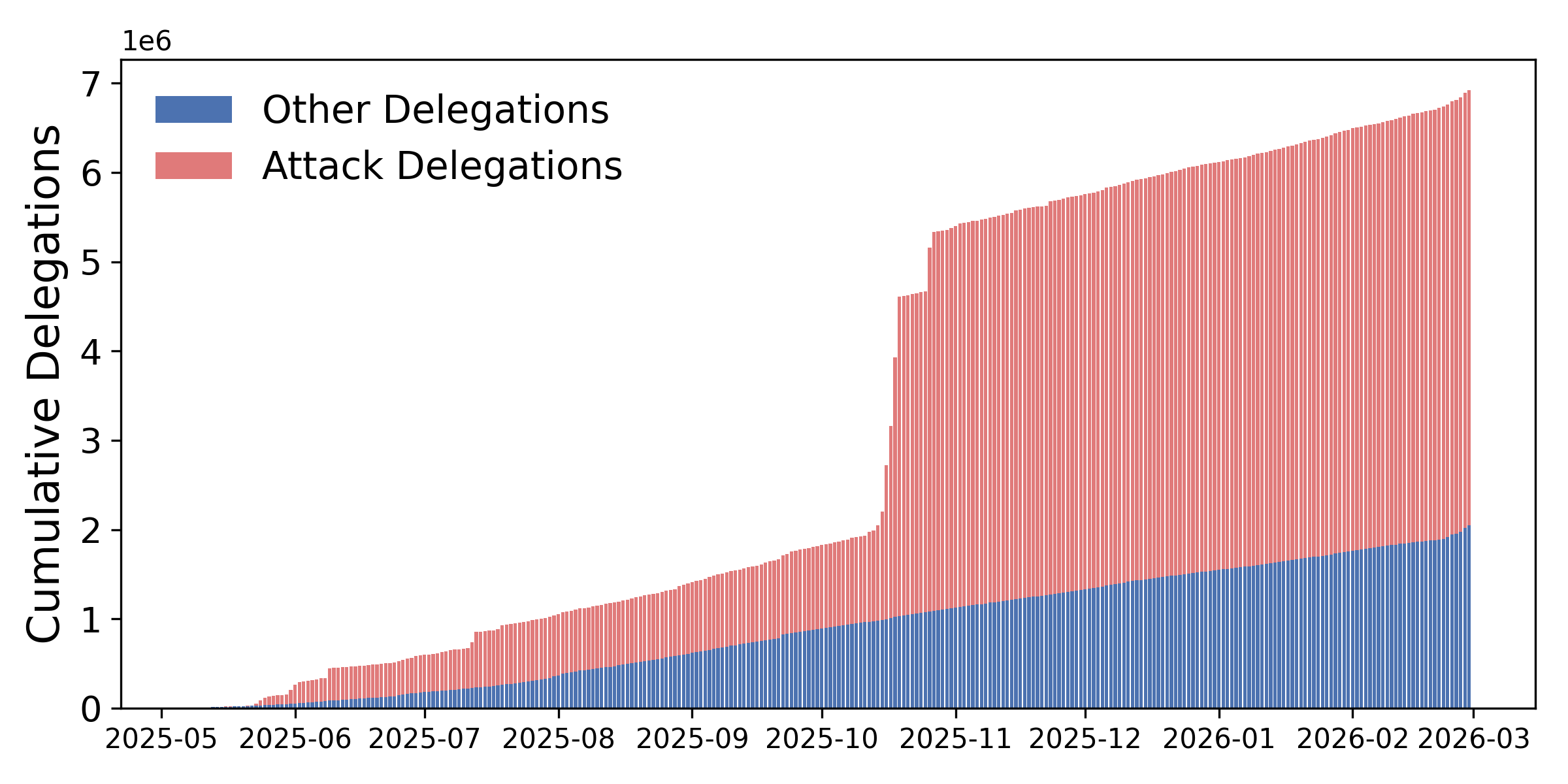}
    \caption{The volumes of malicious and other delegations. Malicious activities constitute a predominant share, reaching 69.88\% of all delegations.}
    \label{fig:malicious_delgations}
\end{figure}

\begin{table}[t]
\centering
\caption{Top 10 Contracts by Number of Delegations and Authorities}
\begin{tabular}{crrc}
\toprule
\textbf{Contract} & \textbf{\# Delegations} & \textbf{\# Authorities} & \textbf{Label} \\
\midrule
\texttt{\href{https://etherscan.io/address/0x072c2673ade3632594337a98a473d32a17505fdb}{0x072c...5fdb}}&	1282,858& 500,339& ETH Sweeper\\
\texttt{\href{https://etherscan.io/address/0x3c3d718877a7d95857e1ae74b3963bddefb81931}{0x3c3d...1931}}&	501,326&	390,633&	ETH Sweeper\\
\texttt{\href{https://etherscan.io/address/0xb59f313dcf8c8107adffeabd0c041c896c64dfca}{0xb59f...dfca}}& 495,951&	244,210&	ETH Sweeper\\
\texttt{\href{https://etherscan.io/address/0xd2e28229f6f2c235e57de2ebc727025a1d0530fb}{0xd2e2...30fb}}&	472,467&	163,619&	Trust Wallet\\
\texttt{\href{https://etherscan.io/address/0x84d05511614272694d3a9cebe896514dbde51f40}{0x84d0...1f40}}&	456,114&	104,079&	ETH Sweeper\\
\texttt{\href{https://etherscan.io/address/0x0e04736a85433445ef602d07946671685ec94647}{0x0e04...4647}}&	435,503&	6,131&	ETH Sweeper\\
\texttt{\href{https://etherscan.io/address/0x8eac07ec5c61138c5472616e39ccb176ca068c25}{0x8eac...8c25}}&	398,336&	302,177&	ETH Sweeper\\
\texttt{\href{https://etherscan.io/address/0x63c0c19a282a1b52b07dd5a65b58948a07dae32b}{0x63c0...e32b}}&	182,205&	175,618&	Metamask\\
\texttt{\href{https://etherscan.io/address/0x5a7fc11397e9a8ad41bf10bf13f22b0a63f96f6d}{0x5a7f...6f6d}}& 177,938&	168,189&	Ambire\\
\texttt{\href{https://etherscan.io/address/0x08f5e230d412f0cd475b03b9acb51d1562f730f9}{0x08f5...30f9}}&	176,176&	169,356&	ETH Sweeper\\
\bottomrule
\end{tabular}
\label{tab:top_contracts}
\end{table}

\textbf{Contract Delegations.} We collect delegations directed toward identified ETH sweeper contracts and define them as malicious delegations. Fig.~\ref{fig:malicious_delgations} illustrates the distribution of malicious and remaining delegations. Out of 6,680,498 total delegations, malicious ones reach 4,668,209, representing 69.88\%. Temporally, while benign delegations grow linearly, malicious ones exploded in October. Specifically, a total of 81,545 malicious delegation transactions occur in October 2025, containing 3,281,091 malicious delegations, which accounts for 70.29\% of all malicious delegations. This highly concentrated phenomenon indicates that attackers are conducting large-scale batch delegation operations. Furthermore, ETH sweeper contracts generally exhibit characteristics of high-frequency delegation. As shown in Table~\ref{tab:top_contracts}, among the top ten contracts with the highest delegation frequencies, seven are ETH sweeper contracts identified by \textit{SweeperHunter}. Notably, more than \textbf{3.74 million} distinct addresses have previously delegated to these seven high-profile malicious contracts. These results not only reflect the rampant nature of attacker behavior but also suggest that an increasing number of attackers are joining this attack ecosystem, highlighting the urgency for early identification and continuous monitoring.

\begin{insightbox}
\textbf{Finding 2}: \textit{Malicious delegations to ETH sweeper contracts account for a staggering 69.88\% of all delegations. We observe a significant surge in such delegations during October 2025, characterized by prominent batch delegation patterns. Notably, 7 out of the top 10 contracts by delegation volume are identified as malicious contracts.}
\end{insightbox}

\textbf{Contract Implementation.} 
Through a systematic manual audit of the 508 contracts in \textit{SweeperDataset}, three experts identify two distinct operational patterns of ETH sweeper contracts based on the underlying redirection mechanisms and reveal several strategic design principles of these contracts.

\begin{figure}[t]
    \centering
    \includegraphics[width=0.7\linewidth]{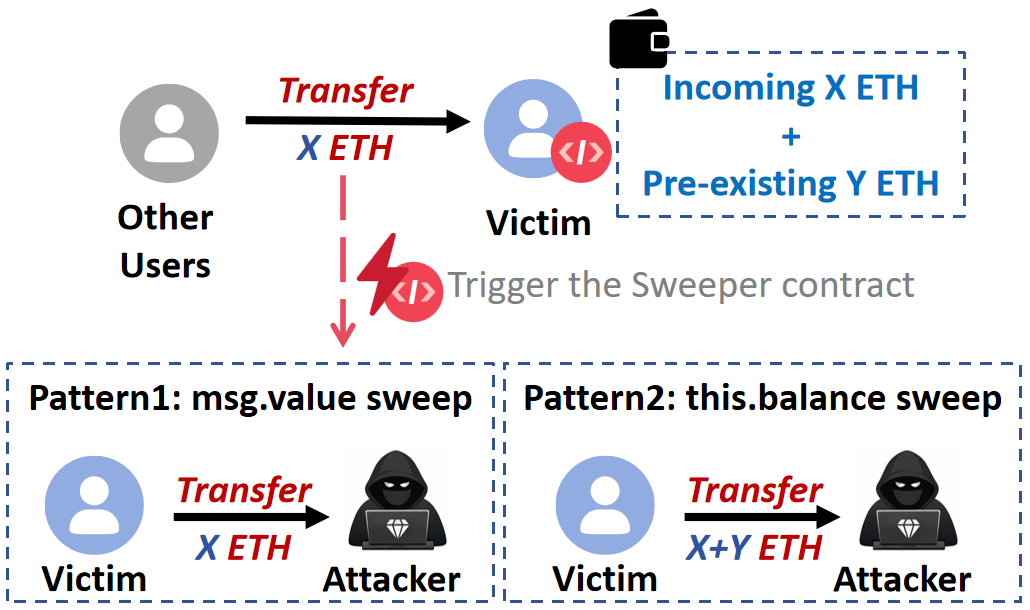}
    \caption{Two patterns of ETH sweeper contracts: \textit{msg.value-based} sweeping redirects funds from third-party transfers, whereas \textit{this.balance-based} sweeping also captures the victim's existing account balance}
    \label{fig: sweep_type}
\end{figure}

\textbf{(1) Redirection Mechanisms.} Based on the scope of the funds being redirected, the two identified patterns are categorized as \textbf{\textit{msg.value-based} sweeping} and \textbf{\textit{this.balance-based} sweeping}. The distinction between these two mechanisms is illustrated in Fig.~\ref{fig: sweep_type}.
The fundamental difference lies in the source of the redirected funds. In an EVM-compatible contract, \texttt{msg.value} represents the specific amount of ETH transferred within the current transaction, whereas \texttt{address(this).balance} refers to the total ETH balance of the account at the time of execution. Within the EERA workflow, the arrival of ETH at the account of the victim triggers the contract code to execute the redirection. Consequently, if a contract transfers the total balance, it redirects both the ETH received in the current transaction and any pre-existing funds of the victim to a destination specified by the attacker. Conversely, the use of \texttt{msg.value} limits the redirection to only the funds originating from third-party accounts within that specific transaction. Analysis reveals that 271 contracts implement the \textit{msg.value} pattern, while 236 contracts adopt the \textit{this.balance} approach.

\textbf{(2) Strategic Design Principles.} We further identify three primary architectural principles that attackers employ to maximize the efficacy and flexibility of the ETH sweeping process:

\begin{itemize}[leftmargin=*]
\item \textit{\textbf{Autonomous Triggering}}. Attackers embed redirection logic into entry points that execute automatically upon receiving ETH. While some contracts use standard hooks such as \texttt{receive} or \texttt{fallback}, others trigger malicious logic through conditional branches when the transaction input is empty. This zero-interaction design enables stealthy and automatic fund redirection during ordinary ETH transfers.
\item \textit{\textbf{Minimalist Design}}. The identified sweeper contracts exhibit a high degree of minimalism, containing only the essential logic for fund redirection. This lightweight design serves a dual purpose: minimizing gas consumption for contract deployment and execution, and reducing code complexity to limit features available for pattern-based detection, thereby ensuring cost-effectiveness at scale.
\item \textit{\textbf{Adaptive Redirection}}. Certain contracts incorporate specific setter functions to dynamically update redirection targets, enabling a flexible operational model. This mechanism allows attackers to dynamically modify the ultimate asset destination without re-submitting an EIP-7702 authorization tuple. Operationally, this capability is strategically deployed to evade reactive, address-based blacklists from exchanges and security vendors. 
\end{itemize}

\begin{insightbox}
    \textbf{Finding 3:} \textit{ETH sweeper contracts typically embed redirection logic in auto-triggering functions, categorized into \textit{msg.value-based} and \textit{this.balance-based} sweeping. Identifying specific code patterns within transfer-related logic constitutes a promising direction for the automated detection of malicious sweeper contracts.}
\end{insightbox}

\subsubsection{\textbf{Profit Destinations}}


\begin{figure}[t]
    \centering
    \subfloat[]{
        \label{fig:ETH_out}
        \includegraphics[width=0.48\linewidth]{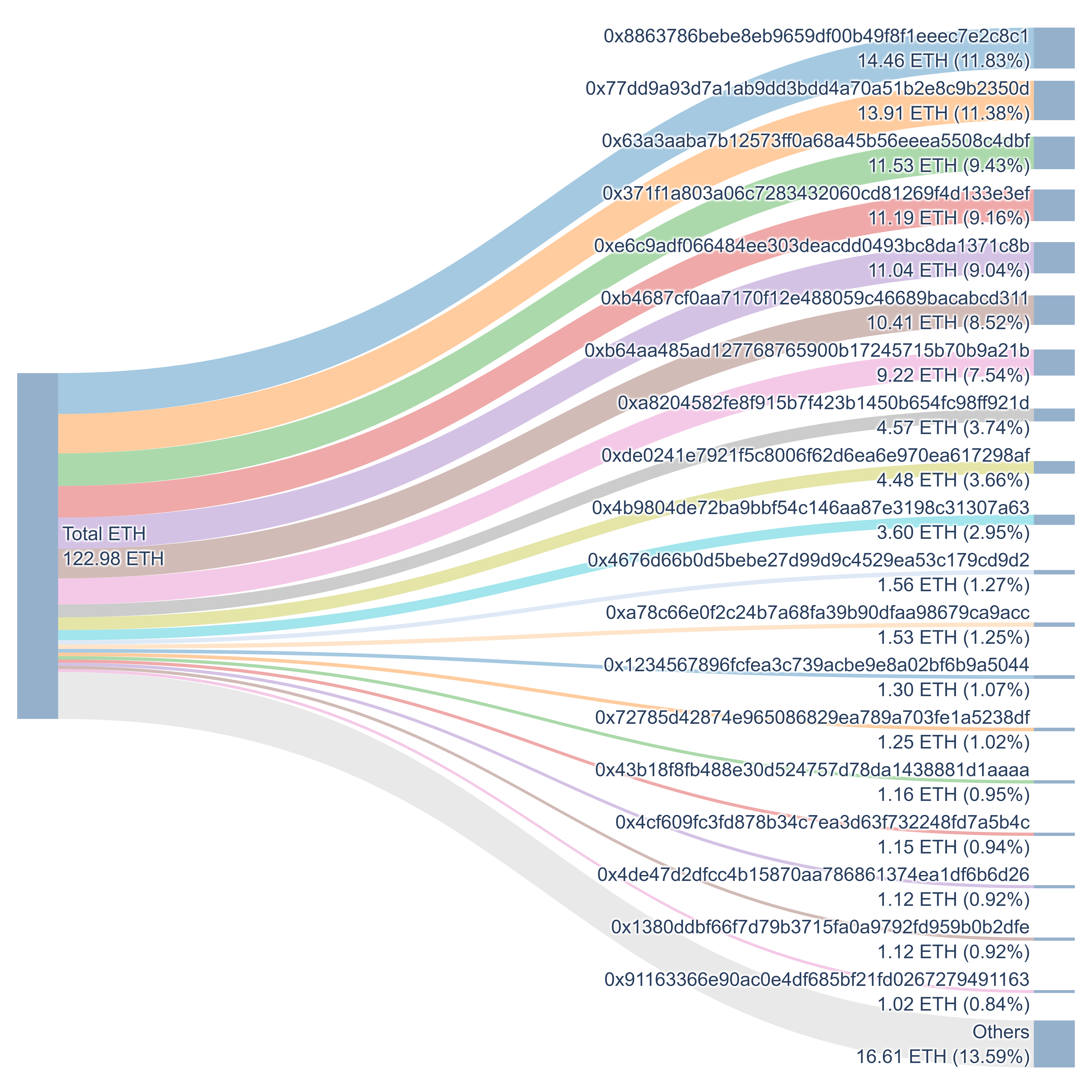}
    }
    \subfloat[]{
        \label{fig:ETH_source}
        \includegraphics[width=0.48\linewidth]{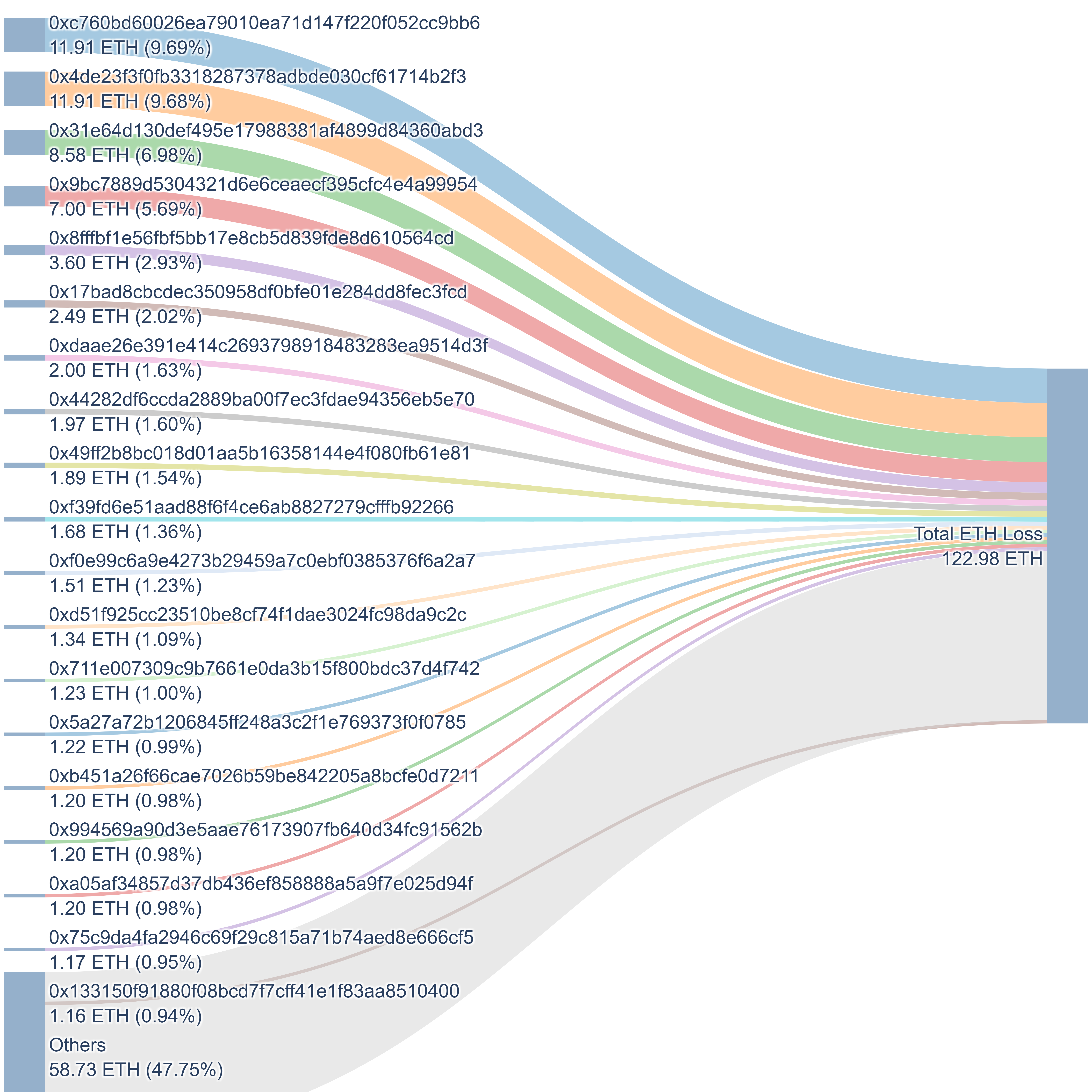}
    }
    \caption{Analysis of swept ETH flows: (a) the destinations of swept ETH, and (b) the sources of swept ETH.}
    \label{fig:ETH_combined}
\end{figure}

We define the addresses to receive stolen funds as \textbf{redirection target addresses}. The Sankey diagram(Fig.~\ref{fig:ETH_out}) illustrating the destinations of the swept ETH. There are a total of 362 distinct redirection target addresses.  Among these, 19 redirection target addresses have swept more than 1 ETH each. The proceeds collected by these 19 target addresses account for over 86.42\% of the total stolen ETH, exhibiting a clear concentration of illicit funds. 5,350 transactions (26.2\%) redirect ETH to the address \texttt{0xe6c9...1c8b}, while 2,235 transactions (10.9\%) redirect ETH to the address \texttt{0x77dd...350d}.

Furthermore, we observe that the relationship between ETH sweeper contracts and redirection target addresses is not a simple one-to-one mapping. On one hand, multiple distinct contracts designate the same fund receiving address. For instance, 40 different ETH sweeper contracts redirect ETH to the address \texttt{0xe6c9...1c8b}. On the other hand, a single contract may correspond to multiple redirection target addresses. This occurs because certain contracts include functions to reset the redirection target address, allowing attackers to invoke the reset function to update the target address.

\subsection{Victims}

\subsubsection{\textbf{Scale of Victims}}


In \textit{SweeperDataset}, a total of \textbf{5,745} distinct victims have already had their ETH swept by attackers. \textbf{607,694} unique accounts have delegated to the detected ETH sweeper contracts, which represent a vast pool of potential victims. The sankey diagram illustrating the sources of swept ETH is shown in Fig.~\ref{fig:ETH_source}. 14 victim accounts experience more than 50 ETH redirection attacks, and 2,172 victim accounts show more than one attack in their transaction history. Notably, the account \href{https://etherscan.io/address/0xc760bd60026ea79010ea71d147f220f052cc9bb6}{\texttt{0xc760...9bb6}} suffers the most significant loss, losing nearly 12 ETH across the four attack transactions, which accounts for 15.10\% of the total losses. Specifically, the primary loss of this victim occurs in an attack transaction in July 2025 (\href{https://etherscan.io/tx/0x073726eaaf9ec62cff58220fc5a4cc07c8bd06961a35b2ff5c52c28ee49680c4}{\texttt{0x0737...80c4}}), where a third-party account \texttt{0x655c...5975} transfers 11.9 ETH to the victim. However, because the victim delegates to the ETH sweeper contract \texttt{0x6B78...E933} at that time, these funds are redirected to the attacker-specified address \texttt{0x8863...28c1}. This incident accounts for the largest individual loss in our dataset and explains the sharp surge in cumulative attacker profit observed in Fig.~\ref{fig:cost_and_profit}.

\textbf{Anomalous Account.} We observe that the account \href{https://etherscan.io/address/0x4DE23f3f0Fb3318287378AdbdE030cf61714b2f3}{\texttt{0x4DE2...b2f3}} (hereafter the \textit{anomalous account}) is involved in 5,735 EERA transactions, accounting for 28.1\% of all detected attacks. We therefore conduct an in-depth analysis of this account. Over a ten-month period, it performs 98,559 delegations, of which 96,745 (98.16\%) target the identified sweeper contracts. The account holds 3,481.63 USDT, but these tokens have been frozen according to the USDT blacklist and therefore cannot be transferred. It has also been repeatedly reported as a gas-trapping scam account~\cite{report1, report2}, with its private key publicly exposed online. Unaware that the USDT is frozen, many users (referred to as \textit{depositors}) transfer ETH to this account as gas fees in an attempt to withdraw the stablecoins. Howerver, the deposited ETH is redirected by malicious delegation contracts. Since the private key is publicly available, any attacker can submit EIP-7702 delegations for the anomalous account, resulting in intense competition to bind it to attacker-controlled sweeper contracts and intercept subsequent deposits. This anomalous account corresponds to a complex scam, demonstrating that EERA can be combined with other attack vectors to facilitate the evolution of attack forms.

\begin{insightbox}
    \textbf{Finding 4:} \textit{EERA has been observed to integrate with existing gas-trapping scams,  upgrading the mechanisms for illicit fund collection. And multiple attackers have vied to control the same high-traffic EOA, frequently updating its delegation to inject their own logic.}
\end{insightbox}

\subsubsection{\textbf{Delegation Characteristics}}

To understand the operational mechanics behind malicious delegations, we investigate their empirical characteristics, focusing on transaction submission patterns and execution structures. Our analysis reveals that malicious delegations exhibit three  characteristics.
\begin{itemize}[leftmargin=*]
    \item \textbf{Passive Submission.} The signing and submission of a 7702 delegation statement are logically decoupled. We define active submission as a behavior where the transaction initiator (\texttt{from\_address}) is identical to the delegating party (\textit{Authority}); otherwise, it is defined as passive submission. We observe a significant asymmetry in the initiation patterns between malicious and other delegations. Specifically, among the identified malicious delegations, the proportion of active submissions is only \textbf{0.000017\%} (81/4,668,209). In contrast, for the remaining delegations, the proportion of active submissions is \textbf{19.19\%} (386,096/2,012,289). Furthermore, among delegations associated with 32 mainstream wallet contracts (e.g., \textit{MetaMask}, \textit{Uniswap}), the active submission rate is \textbf{24.34\%}. This near-zero active initiation rate provides direct, empirical evidence that virtually all malicious delegation transactions are submitted to the blockchain by third-party accounts rather than the victims themselves.

    \item \textbf{Extreme Batching.} The 4,668,209 malicious delegations are realized through only 641,379 transactions, resulting in an average of 7.28 delegations per transaction. In contrast, the remaining delegations average only 1.35 delegations per transaction. Notably, we identify 17,090 highly concentrated transactions, each of which simultaneously delegates at least 50 independent accounts to the same malicious contract (with some bundling up to 1,000 delegations). 

    \item \textbf{Exploitation of Leaked Private Keys.} Another characteristic is the systematic exploitation of historically compromised accounts. By evaluating our dataset against the repository of compromised Ethereum private keys collected by Shao et al.~\cite{lost_in_the_address}, we found that, out of the 4,189 leaked addresses they reported, \textbf{4,081} accounts experienced malicious delegations during our observed periods. These compromised accounts accounted for \textbf{334,045} malicious delegations, with an overwhelming majority (\textbf{85.09\%}) bundled inside batched delegation transactions. This evidence reveals that the bulk exploitation of compromised private keys has already become a prevalent attack vector facilitated by EIP-7702.
\end{itemize}

\begin{insightbox}
    \textbf{Finding 5:} \textit{Malicious delegations exhibit extreme batching patterns driven by third-party execution. Attackers have actively exploited the EIP-7702 mechanism to orchestrate large-scale attacks targeting historically leaked private keys.}
\end{insightbox}

\subsubsection{\textbf{Victim Remediation Measures}}

We further analyze whether victims initiate remediation measures after a malicious delegation. One primary method for a victim to counter EERA is to sign a new delegation transaction, re-delegating the account to a benign contract. Therefore, we analyze the delegation updates of victim accounts following a malicious delegation. We categorize these updates into four types: (1) \textit{Delegation to another known ETH sweeper contract}. For example, the anomalous account mentioned before is targeted by multiple attackers, and its delegations typically update from one malicious contract to another. (2) \textit{Delegation to unlabeled contracts}. Unlabeled contracts refer to contracts not identified as ETH sweeper contracts by \textit{SweeperHunter}. Notably, unlabeled contracts include both benign contracts and potential ETH sweeper contracts. Since \textit{SweeperHunter} only detects contracts involved in happened attacks, those containing sweeping logic but yet to trigger an attack remain unlabeled. (3) \textit{Delegation to the zero address}, which is equivalent to the account executing no code and reverting to a standard EOA. (4) \textit{No update}, meaning no subsequent delegation occurs after a malicious one.

\begin{table}[t]
  \centering
  \caption{Distribution of delegation updates}
  \label{tab:delegation_stats}
  \begin{tabular}{ccccc}
    \toprule
    \textbf{Type} & \makecell[c]{\textbf{Type 1} \\ \small To malicious} & \makecell[c]{\textbf{Type 2} \\ \small To unlabeled} & \makecell[c]{\textbf{Type 3} \\ \small To zero} & \makecell[c]{\textbf{Type 4} \\ \small No update} \\
    \midrule
    \textbf{Quantity} & 3,897,444 & 158,521 & 8,757 & 603,487 \\
    \textbf{Ratio}    & 83.50\%   & 3.40\%  & 0.20\% & 12.90\% \\
    \bottomrule
  \end{tabular}
\end{table}


The statistical results of delegation updates are illustrated in Table~\ref{tab:delegation_stats}. Over 83\% of malicious delegations are followed by new malicious delegations, which further supports the inference that the private keys of these accounts have been compromised, allowing multiple attackers to continuously re-delegate them to different malicious contracts. A total of 607,694 accounts have delegated to detected ETH sweeper contracts. However, only 4,959 authorities, in fewer than two-thousandth (8,57 times) of malicious delegations, re-delegate the accounts to the zero address. More strikingly, 603,487 accounts (99.3\%) do not update their delegations after the final malicious delegation. This implies that as of late October 2025, these accounts remain delegated to ETH sweeper contracts and continue to be at risk.

\begin{insightbox}
    \textbf{Finding 6:} \textit{There is a significant lack of proactive mitigation among victims of EERA. Of the 607,694 compromised accounts, more than 99.3\% remain in a state of risk, as they have not revoked or updated their malicious delegations.}
\end{insightbox}

%% file: sections/discussion.tex
\section{Discussion}\label{sec:discussion}

\subsection{Contribution to the Community}

The prevalence of ETH redirection attacks based on EIP-7702 has surpassed 4.66 million malicious delegations, which establishes this phenomenon as an urgent and evolving attack paradigm. We have reported our detection results and key insights to GoPlus, a prominent Web3 security firm, and received prompt feedback and acknowledgement. It is worth noting that while Goplus has previously disclosed isolated malicious delegation contracts, a systematic and detailed detection or analysis regarding EERA remains absent prior to our work. Currently, we are collaborating with GoPlus to release security advisories to raise awareness within the community and among on-chain users regarding this emerging threat. Additionally, we have submitted the identified ETH sweeper contracts to Chainabuse, a prominent community-driven reporting platform managed by TRM Labs, to facilitate broader awareness and promote the timely labeling of malicious contracts.

\subsection{Attack Impact}\label{discussion_impact}

We emphasize that the losses quantified in this paper represent only a lower bound of the actual impact of EERA, which has already emerged as a serious threat. First, EERA represents a new attack paradigm introduced by EIP-7702. Similar attacks are likely to arise on other EVM-compatible chains that support EIP-7702 or comparable mechanisms, including major platforms such as Optimism, Base, BNB Chain, and Polygon. Similar sweeper contracts and malicious delegation behavior have already been observed on these chains~\cite{PolygonContract}\cite{BNBContract}, while the associated losses are not included in our analysis.
Second, EERA is inherently persistent. As we discuss in Section~\ref{sec:Threat}, there is a key shift in the attacker monetization paradigm. The profit of attackers mainly depends on future incoming transfers from third-party accounts to the authority addresses. Our analysis shows that 98.8\% of affected accounts do not update or revoke delegation after compromise, which leaves them continuously exposed. As a result, the financial impact of EERA will continue to increase over time.

\subsection{Mitigation Recommendations}
\textbf{User-level Mitigation.} Given that private key leakage is the primary attack vector, users must strictly safeguard their credentials and, in particular, avoid entering or pasting private keys and recovery phrases on unverified platforms. Upon realizing a potential compromise, users should immediately transfer remaining assets to a secure address. Furthermore, to mitigate ongoing risks, users must block potential fund inflows by notifying frequent senders to cease transfers.

\textbf{Wallet-level Mitigation.} Wallet providers are expected to implement some new features to mitigate EERA: (i) \textit{Delegation Visualization}: Provide persistent status checks and UI indicators to inform users of active code delegations; (ii) \textit{Dynamic Risk Alerts}: Match delegation targets against blacklists to interrupt high-risk interactions; and (iii) \textit{One-click Revocation}: Integrate a ``Reset to EOA'' feature that automates delegation to the zero address, allowing victims to rapidly neutralize malicious logic.

\textbf{Community-level Mitigation.} We urge security firms to actively monitor EERA and publish reports to bridge the current awareness gap regarding EIP-7702 risks. Moreover, as ETH sweeper contracts are the critical payloads, stakeholders (including blockchain explorers and auditing agencies) should collaborate to disclose a public database of identified malicious sweeper contracts.

%% file: sections/related_work.tex
\section{Related Work}\label{sec:related_work}

\textbf{Analysis of Ethereum Improvement Proposals (EIPs).} Several empirical and security studies have examined evolving EIPs. For EIP-1559, \textit{Leonardos et al.}~\cite{leo1559} and \textit{Liu et al.}~\cite{Liu1559} evaluate its fee stability, user bidding behavior, and consensus security, while \textit{Arshi et al.}~\cite{Arshi1559} demonstrate that block congestion attacks remain feasible. For EIP-4844, \textit{Park et al.}~\cite{park} show that it increases Rollup utilization at the expense of a slightly higher fork rate. Regarding the recent EIP-7702, \textit{Huang et al.}~\cite{7702_usenix} evaluate overall asset exposure and structural vulnerabilities in smart accounts. In contrast, our work zeroes in on a specific, insidious threat vector: a novel, passive attack paradigm termed EERA, which has neither been formally defined nor systematically analyzed.

\textbf{On-chain Attack Detection.} Detection methodologies for on-chain attacks primarily fall into two categories: pattern-based and machine learning-based methods. Pattern-based methods typically rely on manually extracted attack flow and transaction patterns. For example, \textit{Chen et al.}~\cite{payload} identify phishing attacks by analyzing transaction payloads combined with heuristic rules. \textit{Zhou et al.}~\cite{www_key} conduct a measurement study on the misuse of leaked secret keys, systematically uncovering and quantifying large-scale asset theft on public blockchains by linking exposed keys from web sources to on-chain transactions.
Additionally, \textit{Guan et al.}~\cite{guanpoison} and \textit{Tsuchiya et al.}~\cite{poisonusenix} detect address poisoning attacks through prefix and suffix analysis of interaction addresses. \textit{Su et al.}~\cite{dappattack} systematically analyze the attack chains and patterns of DApp attacks to achieve detection, while \textit{Li et al.}~\cite{solphishhunter} define and detect novel phishing attack patterns specifically on the Solana.
Machine learning-based methods leverage labeled data to train models for identifying complex attacks. \textit{Zhang et al.}~\cite{txspector} propose Txspector, using transaction sequence features for on-chain detection. \textit{Chen et al.}~\cite{soda} construct a general online detection framework SODA using machine learning models to identify smart contract attacks. \textit{Wu et al.}~\cite{defiranger} train models based on behavioral features to detect DeFi price manipulation. \textit{Wu et al.}~\cite{zyattack} design a general detection method based on contrastive learning and transaction reconstruction, effective even with zero or few samples of novel attacks. Graph-based methods have also been widely explored. \textit{Wu et al.}~\cite{crossattack} employ graph representation learning to identify cross-chain attacks. \textit{Du et al.}~\cite{tifs_malicious} leverage dynamic graph learning with class-incremental learning for fine-grained malicious account detection in Ethereum. \textit{Jin et al.}~\cite{tifs_fraud} enhance Ethereum fraud detection through generative and contrastive self-supervised learning over transaction graphs. Distinguished from existing works, our work focuses on the delegation execution semantics introduced by EIP-7702, revealing and detecting a novel paradigm EERA. EERA is distinctly different from previously studied attacks in terms of attack patterns and transaction characteristics.

%% file: sections/conclusion.tex
\section{Conclusion}\label{sec:conclusion}

This paper reveals a novel attack vector introduced by EIP-7702, termed the EIP-7702-based ETH Redirection Attack (EERA). In this paradigm, attackers leverage ETH sweeper contracts to ensure that ETH received by victim accounts is automatically redirected to specified addresses. We first delineate the systematic workflow of EERA, based on which we identify 20,443 attack transactions and 508 confirmed ETH sweeper contracts within the on-chain data from May 2025 to February 2026. Then we analyze the behaviors of both attackers and victims based on the identified attacks. Notably, the results show that over 4.66 millions (69.88\%) EIP-7702 delegations are malicious and 607,694 unique accounts have delegated to the identified ETH sweeper contracts, which underscores the prevalence of such illicit activities. 
Furthermore, our analysis demonstrates that the systematic exploitation of historically leaked private keys serves as a critical vector for malicious delegations, enabling attackers to  orchestrate large-scale attacks against compromised accounts.
Finally, we provide targeted mitigation strategies and have shared our findings with the community, receiving positive feedback. This work is expected to provide novel insights into the ongoing efforts in attack prevention.